\documentclass[superscriptaddress, preprintnumbers,amsmath,amssymb]{revtex4}

\usepackage{graphicx}
\usepackage{bm}
\include{epsf}

\begin{document}

\preprint{ preprint for  Entropy }

\title{\Large \bf France new regions  planning? Better order or more disorder ?  }
\author{   Marcel Ausloos$^{1,2,3,*}$  }  
 \affiliation{ $^1$
School of Management, University of Leicester, 
 University Road, Leicester, LE1 7RH, UK\\$^{2}$eHumanities
group\footnote{Associate Researcher}, Royal Netherlands
Academy of Arts and Sciences, \\  Joan Muyskenweg 25, 1096 CJ
Amsterdam, The Netherlands \\  $^3$GRAPES\footnote{Group
of Researchers for Applications of Physics in Economy and Sociology},
 rue de la Belle Jardiniere 483,     B-4031, Angleur, Belgium \\$e$-$mail$ $address$:
marcel.ausloos@ulg.ac.be 
} %

  \date{\today}
\begin{abstract}
This paper grounds the critique of the 'reduction of regions in a country'  not only in its geographical and social  context but also in its entropic space. The various recent plans leading to the reduction of the number of regions in metropolitan France  are discussed,  based on the mere distribution in the number of cities in  the plans and analyzed according to various distribution laws. Each case, except the present   distribution with 22 regions, on the mainland,  does not seem to fit presently used theoretical models. Beside,  the number of inhabitants is examined in each plan. The same conclusion holds. Therefore a theoretical argument based on entropy considerations is proposed, thereby pointing to whether  more order or less disorder is the key question, - discounting political considerations. 
\end{abstract}

    \maketitle
    
        \vskip0.5cm

 {\bf Keywords: }    \vskip0.2cm
 France; regional planning; number of inhabitants; number of cities; entropy
considerations 

\section{ Introduction}\label{Introduction}

\underline{According to  United Nations  Department of Economic and Social Affairs reports \cite{ONU_2009}, }the majority of the world  population
resides  in urban areas for the first time in
human history. Municipalities  are emerging as key sites
of social experimentation and problems
in the 21st century \cite{Bettencourt and West_2010,Glaeser_2011,Grabar_2013,Katz and Bradley_2013}. In several   countries, the number of municipalities  were or will be  regrouped to form larger entities, "cities",  supposedly more manageable in various ways. Obviously, the number of municipalities  will be smaller.  There is a tendency  to see the  number of municipalities   as a kind of universal, rational and  officially depoliticised process, since bearing on numbers rather than population size or wealth, in contrast to and in order to avoid claims of gerrymandering \cite{Martis_2008}. Both geographers and planners have been
using increasingly sophisticated quantitative
and computational methods to explain, suggest or even recommend  the number reduction  \cite{Cox and Mair_1988,McCann and  Ward_2010}. The point is not discuss  such hypotheses  nor "reasonings" which are, as usually admitted, quite debatable, and surely time dependent.  Interestingly, another type of  number reductionism process occurs at the next  administrative level, i.e. the reduction in the number of departments, provinces, regions, depending on the country structure, \underline{e.g. as can be  illustrated through European statistics  \cite{NUTS_2013}. }

Consider France (FR), as a timely example. The country  is made of 27 regions and 101 departments: 22 regions constitute the "Metropolitan France",  while 5 regions, which are also departments, make the Overseas Territories, called DOM-TOM.   (DOM: D\'epartements d'Outre-Mer, or overseas departments; TOM = "Territoires d'Outre-Mer", or overseas territories. Both are often grouped as  DOM-TOM.)  Several proposals  have been presented about   regrouping the departments into a smaller number of new regions, - supposedly  for various administrative and economic reasons, mainly,  i.e. to reduce administrative spending, as it has been claimed, but also surely according to some public opinion, to concentrate political power. Some considered regrouping,  from 22 to 15, $\dots$,  down to 11 regions, envisaged at various recent times, is briefly recalled here below in Sect. \ref{chronology}.   On Wednesday  Dec. 17, 2014, the French Parliament voted to  reduce the number of regions to  13  with  still some possibility for departments to negotiate their regional membership later.  (The reform will be in effect in   2016). There were and still are many manifestations of anger concerning the new regional distribution and content, for diverse reasons. This anger is not of real concern for this paper. Nevertheless, an objective data analysis might shine some light on the process.

The number of departments will remain unchanged.  The number of elected councilors will also remain the same.  Of course, political power might be modified according to the assembly member political affiliations.  The criterion has not been mentioned, although in a related affair, i.e. the regrouping of municipalities  in Belgium,  to redistribute local power was the main criterion.  Nevertheless, let us neglect such a point of view at this level.

Thus, neglecting political constraints, it seems of interest to observe whether the regrouping of regions in FR obeys theoretical laws on the grouping of settlements in  an area, according to Gibrat  
 \cite{Gibrat} or 
or Yule  \cite{yule1922}
 models and subsequent laws. These are not universally obeyed, but  they are based on scientific reasoning tied to common sense    \cite{BGNEGNKVZID}. 
In that spirit, it is fair to mention a broader  study of the parametric description of city size distributions (in  four
European countries: France, Germany, Italy and Spain) by  Puente-Ajovin and Ramos \cite{Puente-Ajovin and Ramos_2015} through several  parametric models.  The results are quite coherent.   Therefore, for our endeavor, there is no need here to investigate the flurry of models as discussed by Puente-Ajovin and Ramos \cite{Puente-Ajovin and Ramos_2015}.

To  test those models or rules, the  standard way of presenting the administrative content of regions  is through the rank-size relationship   \cite{BGNEGNKVZID,GabaixIoannides04,Brakmanetal99,ZDMA},
 i.e. a display of the list of regions ranked from $r=1$ to its maximum value $r_M$, i.e. the number of  regions, $N_r$. The ranking is based on the number of municipalities  in each  region $N_{c,r}$,  given in decreasing order. For example,   Midi-Pyr\' en\' ees is the French region having the largest number of municipalities  (3020), in recent times. The  present list is given in Table \ref{TableNcityperregionFR}.

 The FR number of municipalities , more than 36 000,  has been varying almost monthly, although not drastically.  Thus,  to take such a number evolution over the last few years  in the following analysis is not mandatory:  the Dec. 2014 law on region regrouping is based indeed on a rather so called stable situation.  Thus, the number of municipalities   which is considered  is that  available in Jan. 2014, when the "final" discussion on the number of regions  and their extent  last line of discussion started in the French National Assembly.  Nevertheless, for convincing the reader, the data on the number of municipalities  distribution is also given in Table \ref{TableNcityperregionFR} for the Jan. 2012 time. 

It will be of common sense  interest to compare considerations on municipalities  with the number of inhabitants,  which is known in contrast to be a very fluctuating quantity. It was above 63 millions for the metropolitan FR in Jan. 2014 according to INSEE official  data   \cite{INSEEdataNib}.    
Thus, beside the (distribution of the) number of municipalities , the (distribution of the) number of inhabitants is examined in each case here below, within the same statistical  framework, along  a    statistical distribution  pertaining to the matter, in Sect. \ref{dataanalysis}. N.B.  The 6 municipalities  without inhabitants in metropolitan FR have been taken all into account.     

Thus,    region ranking will be found through their  $N_{i,r}$  content. It should be understood at once that the ranking   $N_{c,r}$  and $N_{i,r}$,  are {\it a priori} not correlated. In fact, only the two highest ranked regions, Limousin and Corse,  maintain their  identical rank  (21 and 22) in both   the number of municipalities  and the number of inhabitants.

It is re-stressed that the paper emphasis is on the region number and city number distribution content as  a basis for estimating the interest of the proposed plans on a purely numerical basis.  Two warnings are  still needed at this point. (i) Some analysis difficulty is  due to Corse. The island   actually forms a region, - with  360 municipalities  all together. This number of municipalities  is clearly an outlier from a statistical point of view.  However, it  will be seen that  political plans have  sometimes  imagined that Corse could be merged with  some other region from Southern France, - see Sect.\ref{chronology}.   
Thus,   for coherence, Corse, although an outlier, has been kept as such in the following data analysis, {\it a priori} knowing  as a consequence that  it would induce a larger error bar on the fit parameters than otherwise.  (ii) The DOM-TOM regions or departments are not included much in the following discussion since the  discussed region regrouping is irrelevant in their case.

A discussion of the fit features is found in Sect. \ref{discussion}. It will appear that the various intended regroupings are not theoretically highly  convincing,  to say the least. A suggestion for a complementary approach will be emphasized with a "thermodynamic/information entropy" argument in Sect.\ref{sec:entropy}. A conclusion is found in Sect.\ref{conclusions}.

Moreover, for some "completeness",  some brief analysis of the  ranking relationship of departments  $only$ according to their number of municipalities    is presented in an Appendix. In so doing, the emphasis on outlier effects, for this case, is quite well seen.

\section{Brief review of a few regrouping plans in a chronological way}\label{chronology} 





The reduction of the metropolitan regions is not a new idea. In 2009, a Committee of reform of the local government agencies, chaired by the former Prime Minister   E. Balladur,   had  one of its proposals, hereby called B15,  recommending  to reduce the number of      regions to 15 areas: Alsace-Lorraine, Auvergne-Limousin, Bourgogne-Franche Comt\' e, Bretagne, Champagne, Corse, Ile de France, Languedoc-Roussillon, Midi-Pyr\'en\'ees, Nord-Pas de Calais, Normandie, Poitou-Aquitaine,  Provence-Rh\^one-Alpes,  Val de Loire-Centre. N.B.  Fusions of the areas are best seen when reading Table  \ref{TableNcityperregionFR}. Notice that  the french names are kept.
  A few names  could be used either  in english or in french:  Burgundy$\leftrightarrow$ Bourgogne, Brittany $\leftrightarrow$  Bretagne,   Corsica $\leftrightarrow$  Corse, Normandy $\leftrightarrow$   Normandie,  Center $\leftrightarrow$  Centre.

In January 2014, President  F. Hollande wished to divide by 2 the number of regions, leading to a map  hereby called M11.   It can be  imagined which would be the list of these 11 new areas:  
 Alsace-Lorraine,  Aquitaine,  Auvergne-Rh\^one, Bourgogne-Franche Comt\'e, Bretagne, 
Grand Paris,  Languedoc-Roussillon,  Nord-Pas de Calais, Normandie, Provence-Alpes-C\^ote  d'Azur-Corse, Val de Loire. N.B. This plan intends to merge  Corse with  the Provence-Alpes-C\^ote  d'Azur region.

However, the Prime Minister M. Valls proposed a territorial  reform on Tuesday, April 8, 2014 based on 12 metropolitan regions, called V12 here below:  
 Aquitaine-Poitou-Limousin,  Artois-Picardie, Bourgogne-Franche Comt\' e, Bretagne, 
Champagne-Alsace-Lorraine-Ardennes,  Corse,  Ile de France, Midi-Pyr\'en\'ees-Languedoc-Roussillon, Normandie,  Provence-Alpes-C\^ote  d'Azur,  Rh\^one-Alpes-Auvergne, Val de Loire.

 Next on Monday, June 2, 2014, F. Hollande  proposed   a map  with 14 new areas, called H14 here below, i.e. Alsace-Lorraine,  Aquitaine,  Auvergne-Rh\^one-Alpes,  Bourgogne-Franche Comt\'e, Bretagne,  Corse, Ile de France,   Midi-Pyr\'en\'ees-Languedoc-Roussillon, Nord-Pas de Calais-Champagne, Normandie, Pays de la Loire, Picardie-Champagne-Ardennes, Poitou-Charente-Limousin-Centre, 
Provence--Alpes-C\^ote  d'Azur.

{\it In fine}, the national Assembly adopted on Friday, July 18, 2014 a   map of France with 13 metropolitan regions.
Due to much discussion with the Senate and local authorities, another  13 new region map  of France was adopted at the national Assembly, in the night of Wednesday 19th November 2014, but finally on Wednesday  Dec. 17, 2014,  the "final" map was approved,     to  be in effect on the horizon 2016. The new  map strictly merges regions, in contrast to other plans which allowed  changes in present region borders. The new regions are: (i) Poitou-Charentes, the Limousin and Aquitaine; (ii) Nord-Pas-de-Calais and Picardy; (iii) Champagne-Ardenne, Alsace and Lorraine; (iv)
Auvergne and Rh\^one-Alpes; (v) Bourgogne and Franche-Comt\'e; (vi)  Languedoc-Roussillon and Midi-Pyr\'en\'ees; (vii) High-Normandie and Lower Normandie.
Six areas  remain unchanged: Brittany,  Corse,  Ile-de-France,  Centre,  Pays de la Loire, and  Provence-Alpes-C\^ote  d'Azur. This will be called the P13 plan.

\section{Data analysis}\label{dataanalysis}
\begin{figure}
\includegraphics[height=7.0cm,width=8.4cm] 
{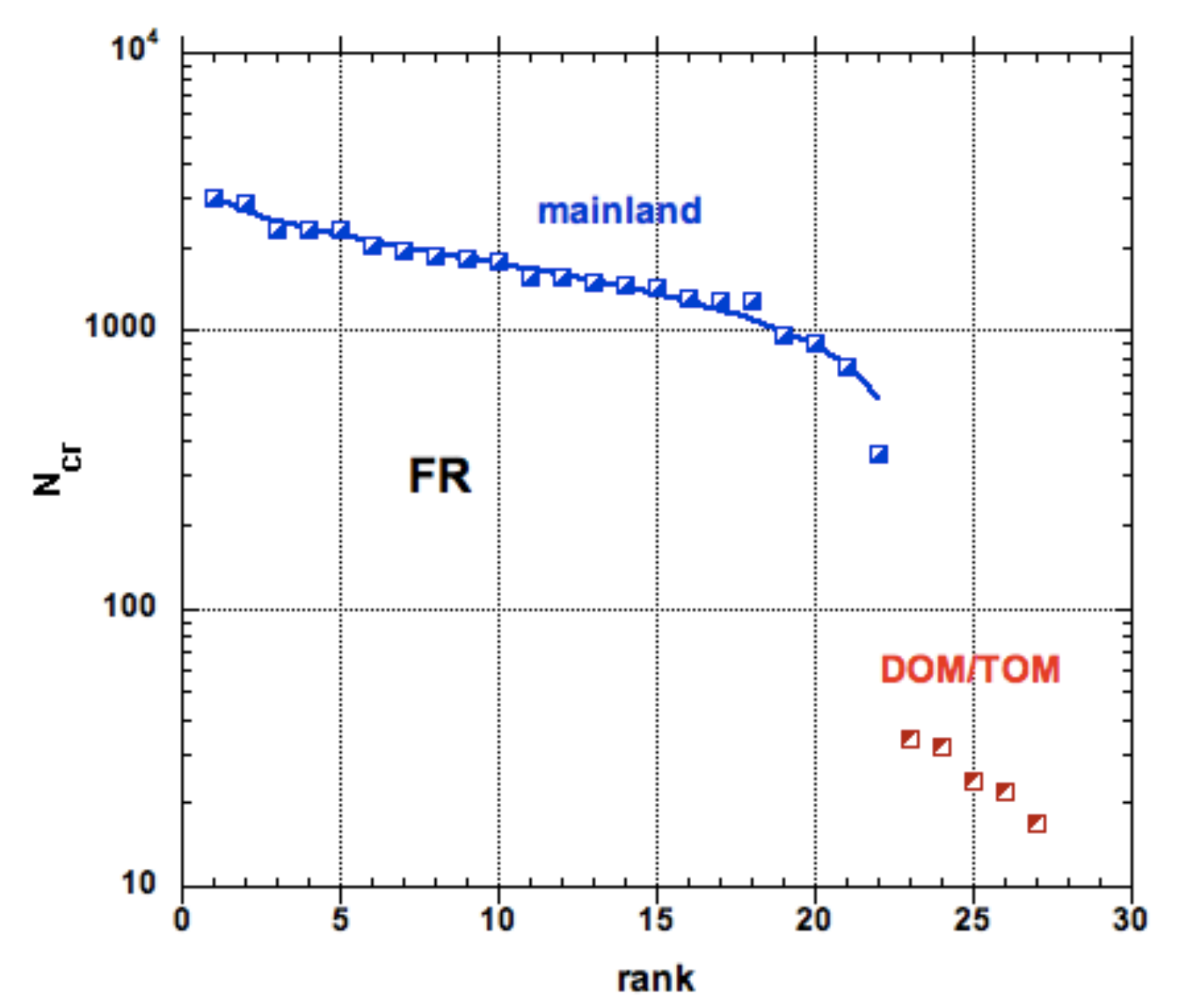}
  \caption   {
  Semi-log plot of the number, $N_{c,r}$, of municipalities  in the 27 FR regions ranked by decreasing order of "importance": blue half filled squares are for the metropolitan area; red half filled squares for the DOM-TOM. The best  3-parameter  function,   Eq. (\ref{Lav3}), fit is   shown  for the metropolitan area only  ($R^2$= 0.978),  but including Corse;  the DOM-TOM have to be considered as outliers.}
 \label{fig:Plot4FRNcrmnlndliloLav3}
\end{figure}

, 

 The present (Jan. 2014) number of municipalities  $N_{c,r}$ in each  of the 27 regions is given in Table \ref{TableNcityperregionFR}, thus including the DOM-TOM for completeness. 
  The rank-size relationship is displayed in Fig. \ref{fig:Plot4FRNcrmnlndliloLav3}.  A fit is proposed through   a reasonable distribution function describing a  rank-size rule  with  3 parameters 
  \begin{equation}
 y(r)= A \; m_1\; r^{-m_2}\;  (N-r+1)^{m_3},
 \label{Lav3} \end{equation}
 where  $A$ is an order of magnitude amplitude, {\it a priori imposed and adapted to the data, without loss of generality, for smoother convergence of the non-linear fit process in finding  the $m_i$ parameters in  a similar order of magnitude range}, and $N$ is the number of regions, of course   ranked, in decreasing order of magnitude of the relevant variable, $r=1, \dots,  r_M$, i.e. the maximum number of  regions, $N_r$. The function has a Beta-Euler function type  \cite{AbraSteg,GradRyz} 
 which has been shown to be useful under the form
  \begin{equation} \label{Lavalette3a}
\kappa_3\;  \frac{(N\;r)^{- \gamma}}  { (N-r+1)^{-\xi}  }, 
\end{equation}
in other related contexts \cite{RRPh49.97.3popescu,Glottom6.03.83popescu,JoI1.07.155Mansilla,JQL18.11.274Voloshynovska,MAJAQM,MALavPRE}. 
 It can be shown that it is related to a
          power law with exponential cut-off,   the so called
 the  Yule-Simon distribution    \cite{Pwco3}
  \begin{equation} \label{PWLwithcutoff}
 y(r)= d \;r^{-\alpha} \; e^{-\lambda r},
\end{equation} 
which is often used in city size distribution studies. However, the latter distribution assumes that the number of elements to be ranked can extend to infinity, while  Eq.(\ref{Lav3}) emphasizes a finite size limit in ranking the data.  In Fig. \ref{fig:Plot4FRNcrmnlndliloLav3}), the data seems well represented by Eq.(\ref{Lav3}), when the 5 (DOM-TOM) outliers are neglected.

For completeness,  a summary of  the statistical characteristics for the number distribution of municipalities   ($N_c $),   in the metropolitan, including Corse,    regions   ($N_{c,r}$) is given in   Table  \ref{Tablestatcityperregiondept12-14FR}. Observe that the number of municipalities   distributions correspond to negative skewness and negative kurtosis.  As expected,the ratio $\mu/\sigma$ is weakly varying, allowing us  for some possible inconsistency through the INSEE data tables. The same holds true the asymmetry of the distributions whatever definition of skewness is used.

\section{Result discussion}\label{discussion}
   
\begin{figure}
\includegraphics[height=7.0cm,width=8.4cm] 
{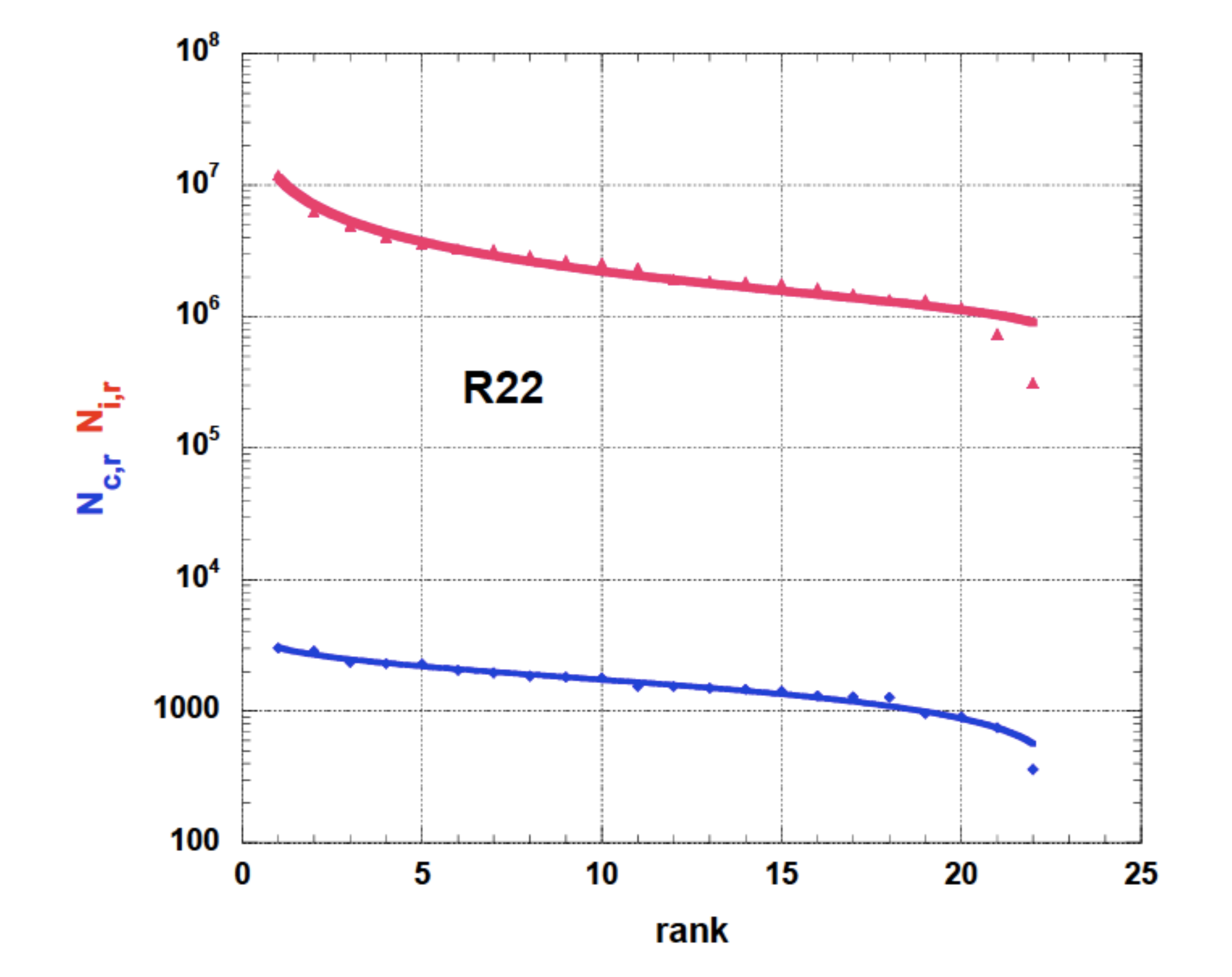}
  \caption   {
  Semi-log plot of the number, $N_{c,r}$, of municipalities  (blue diamonds) and  $N_{i,r}$, of inhabitants (red triangles), in the 22 FR mainland regions ranked by decreasing order of  importance for the R22 plan. The best  3-parameter  function,   Eq. (\ref{Lav3}), fit is   shown for values and regression coefficient   found in Table \ref{table:parametersFRLav3}.}
 \label{fig:Plot22FRNcrNirliloLav3}
\end{figure}

\begin{figure}
\includegraphics[height=7.0cm,width=8.4cm] 
{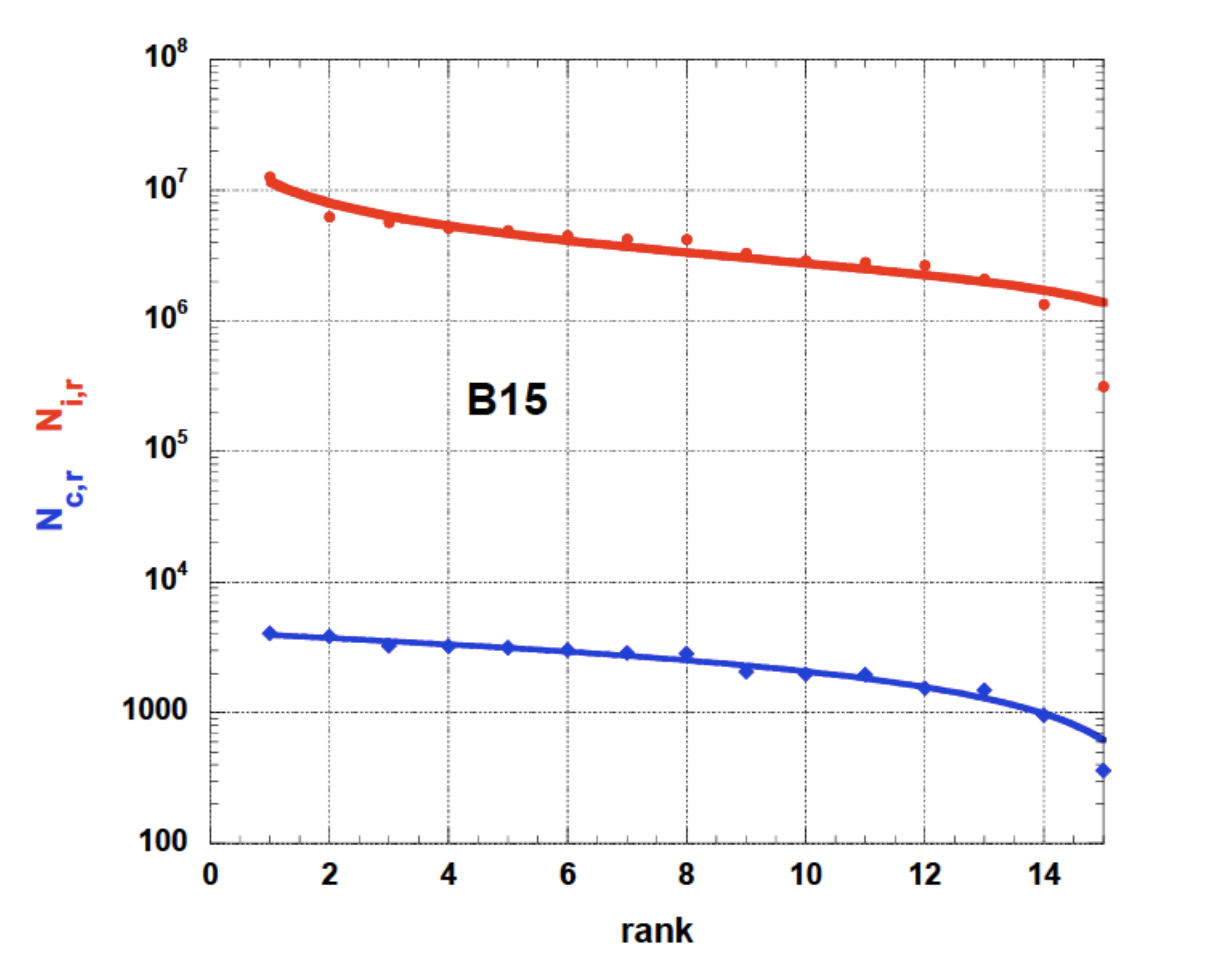}
  \caption   {
  Semi-log plot of the number, $N_{c,r}$, of municipalities  (blue diamonds) and  $N_{i,r}$, of inhabitants (red triangles), in the 22 FR mainland regions ranked by decreasing order of  importance for the B15 plan. The best  3-parameter  function,   Eq. (\ref{Lav3}), fit is   shown for values and regression coefficient   found in Table \ref{table:parametersFRLav3}.}
 \label{fig:Plot15FRNcrNirliloLav3}
\end{figure}

\begin{figure}
\includegraphics[height=7.0cm,width=8.4cm] 
{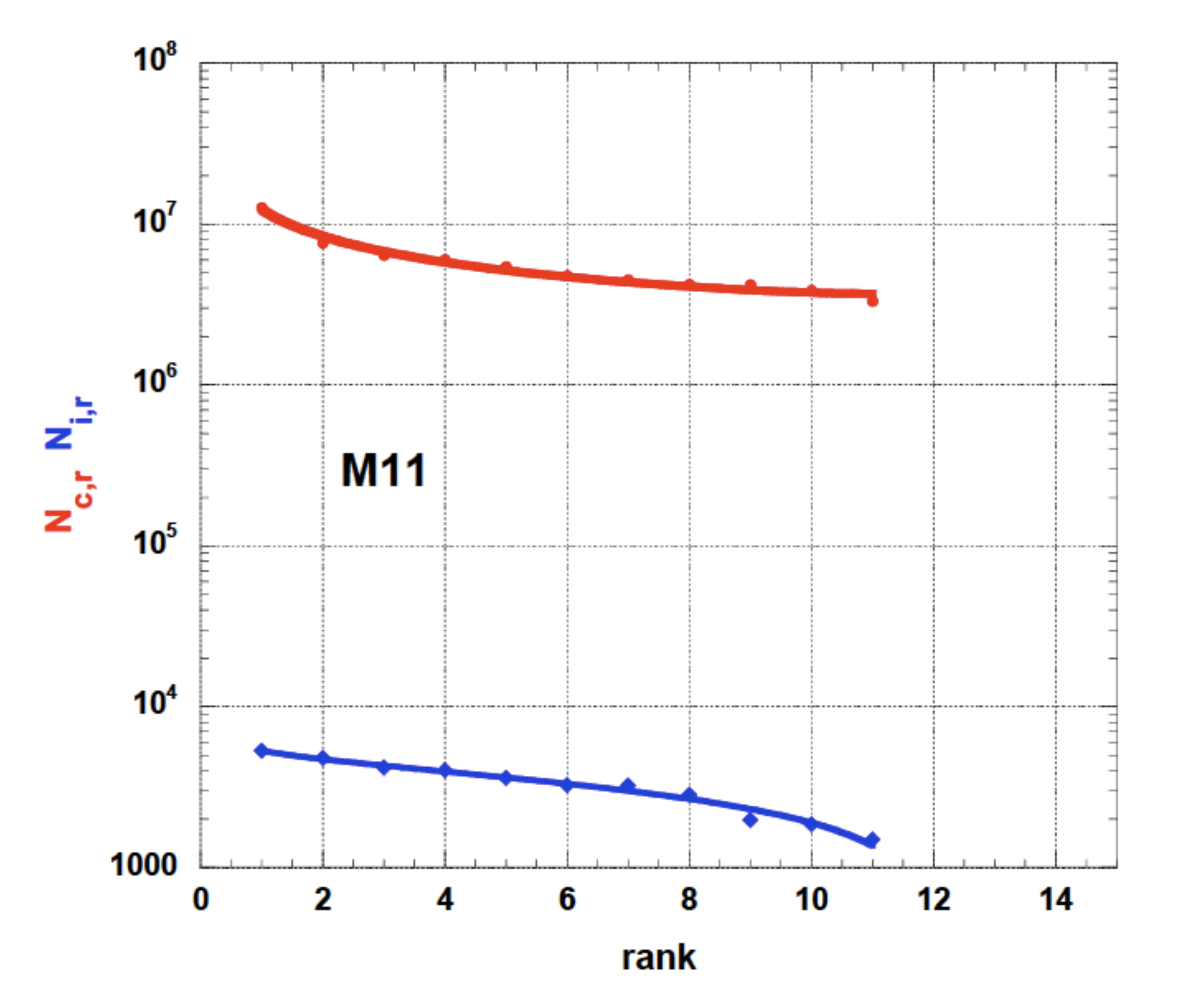}
  \caption   {
  Semi-log plot of the number, $N_{c,r}$, of municipalities  (blue diamonds) and  $N_{i,r}$, of inhabitants (red triangles), in the 22 FR mainland regions ranked by decreasing order of  importance for the M11 plan. The best  3-parameter  function,   Eq. (\ref{Lav3}), fit is   shown for values and regression coefficient   found in Table \ref{table:parametersFRLav3}.}
 \label{fig:Plot11FRNcrNirliloLav3}
\end{figure}

\begin{figure}
\includegraphics[height=7.0cm,width=8.4cm] 
{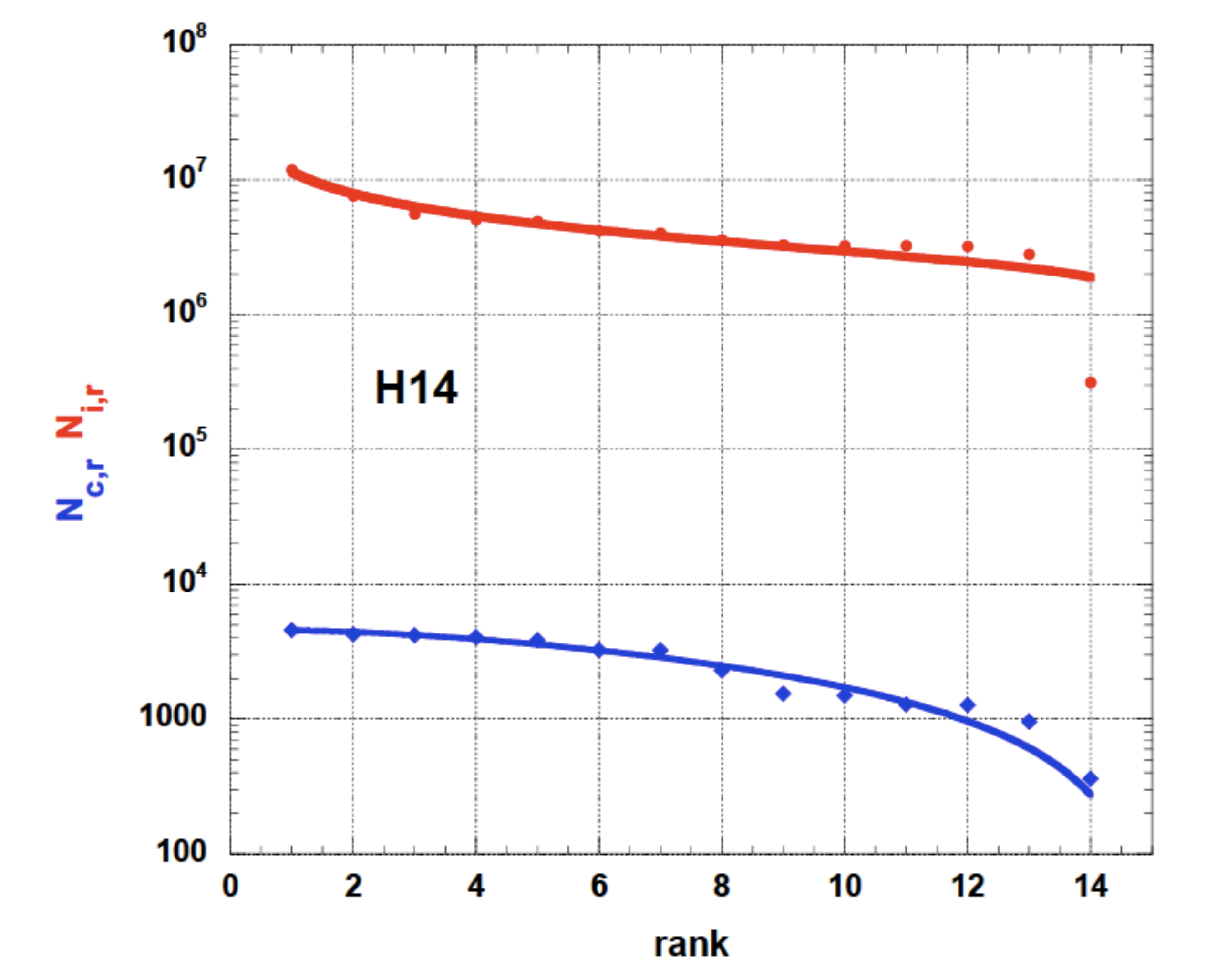}
  \caption   {
  Semi-log plot of the number, $N_{c,r}$, of municipalities  (blue diamonds) and  $N_{i,r}$, of inhabitants (red triangles), in the 22 FR mainland regions ranked by decreasing order of  importance for the H14 plan. The best  3-parameter  function,   Eq. (\ref{Lav3}), fit is   shown for values and regression coefficient   found in Table \ref{table:parametersFRLav3}.}
 \label{fig:Plot14FRNcrNirliloLav3}
\end{figure}

\begin{figure}
\includegraphics[height=7.0cm,width=8.4cm] 
{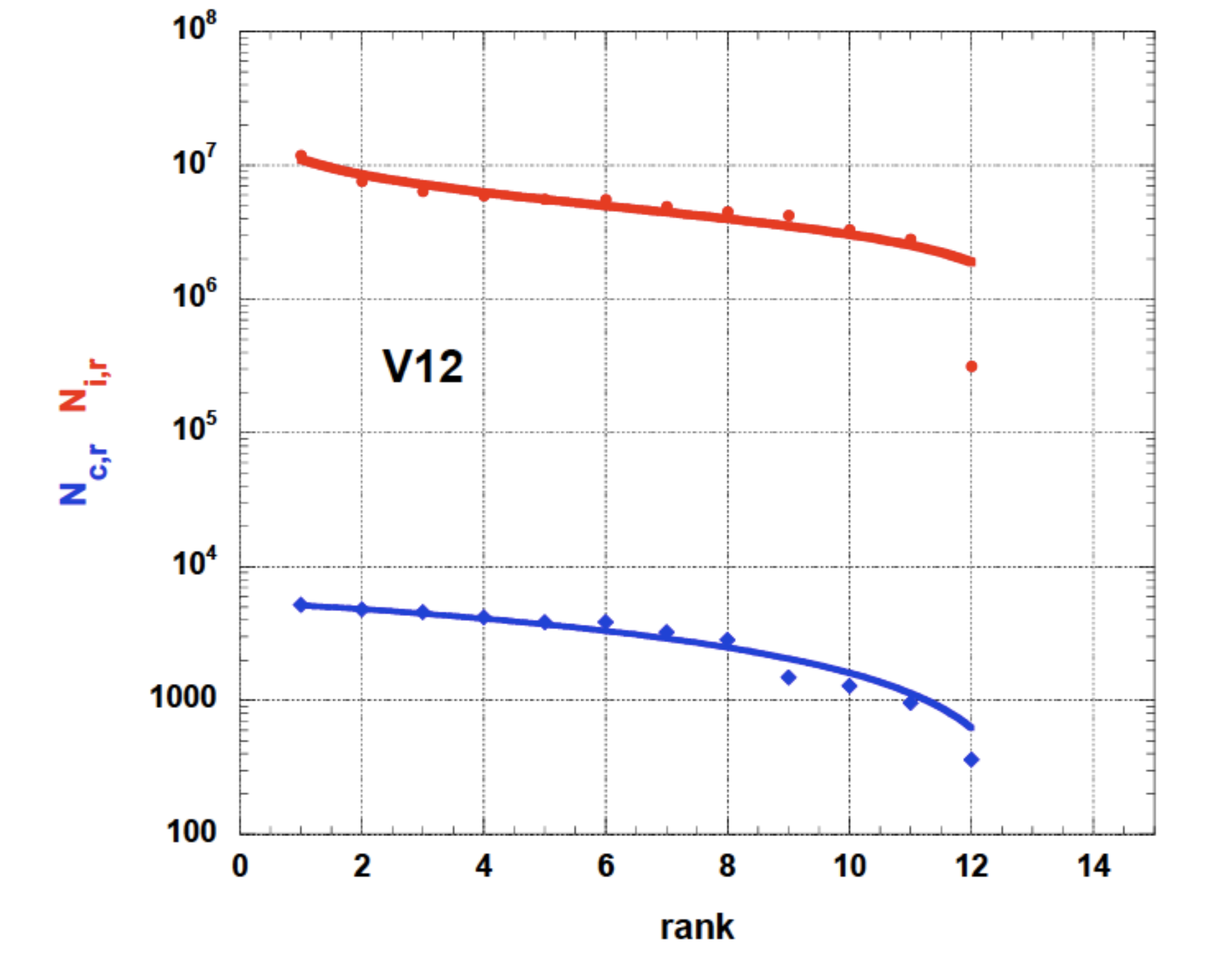}
  \caption   {
  Semi-log plot of the number, $N_{c,r}$, of municipalities  (blue diamonds) and  $N_{i,r}$, of inhabitants (red triangles), in the 22 FR mainland regions ranked by decreasing order of  importance for the V12 plan. The best  3-parameter  function,   Eq. (\ref{Lav3}), fit is   shown for values and regression coefficient   found in Table \ref{table:parametersFRLav3}.}
 \label{fig:Plot12FRNcrNirliloLav3}
\end{figure}

\begin{figure}
\includegraphics[height=7.0cm,width=8.4cm] 
{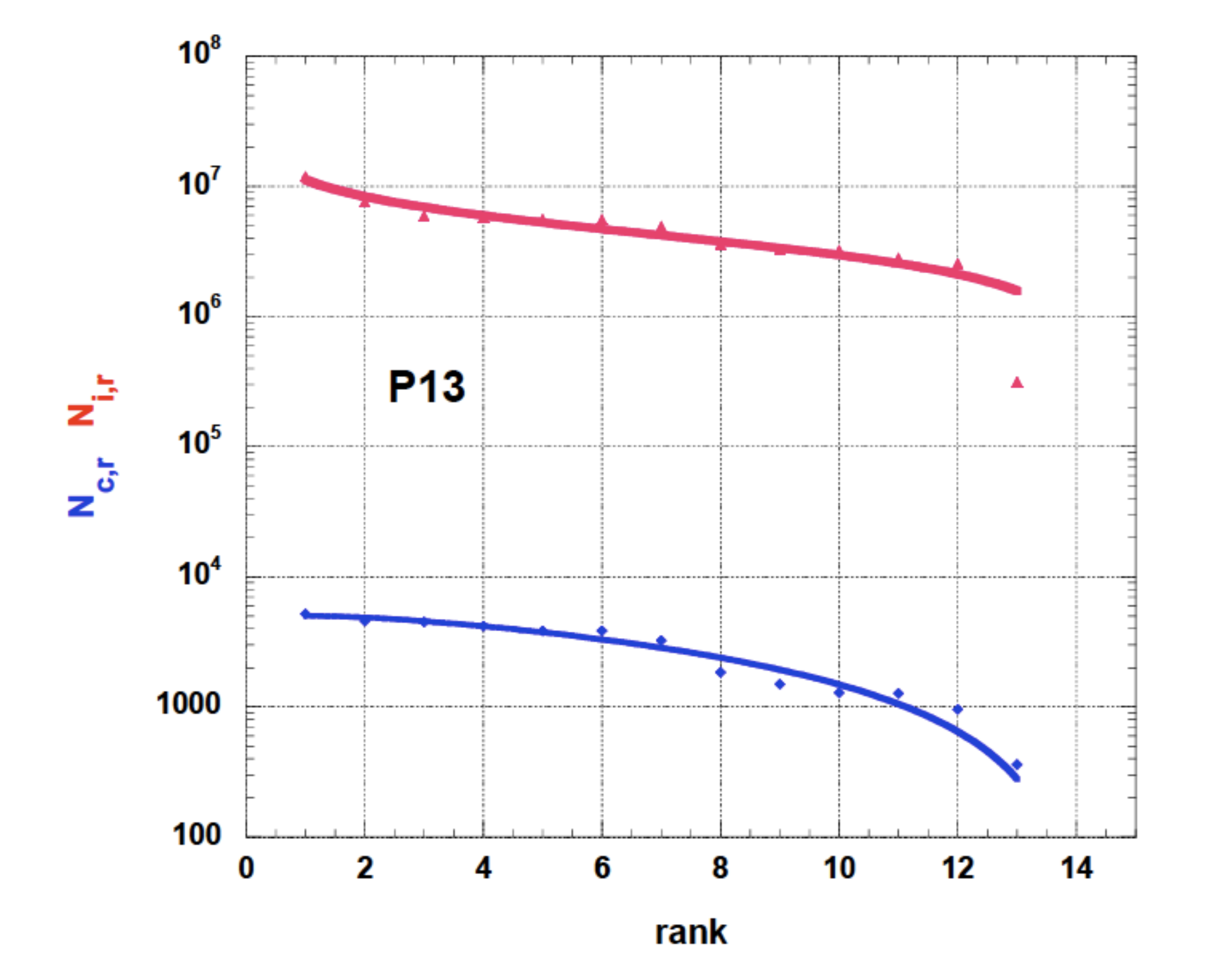}
  \caption   {
  Semi-log plot of the number, $N_{c,r}$, of municipalities  (blue diamonds) and  $N_{i,r}$, of inhabitants (red triangles), in the 22 FR mainland regions ranked by decreasing order of  importance for the P13 plan. The best  3-parameter  function,   Eq. (\ref{Lav3}), fit is   shown for values and regression coefficient   found in Table \ref{table:parametersFRLav3}.}
 \label{fig:Plot13FRNcrNirliloLav3}
\end{figure}

  Let the main attention be focussed on the  region reduction number plans for metropolitan FR.   Consider the various $N_{c,r}$, 	and also the various possible number of inhabitants $N_{i,r}$ within each plan.  The $N_{c,r}$ data of interest for continental FR   is compared to the behavior of  the corresponding  $N_{i,r}$ data  for the present 22 regions on  Fig. \ref{fig:Plot22FRNcrNirliloLav3}.
 The successive figures, Fig. \ref{fig:Plot15FRNcrNirliloLav3}, 
Fig. \ref{fig:Plot11FRNcrNirliloLav3},
Fig. \ref{fig:Plot14FRNcrNirliloLav3},
Fig. \ref{fig:Plot12FRNcrNirliloLav3},
Fig. \ref{fig:Plot13FRNcrNirliloLav3}
 display the corresponding data in the various cases, according to the mentioned chronology of  the successive plans, presented in Sect. \ref{chronology}. The fit parameters are given in Table \ref{table:parametersFRLav3} with the corresponding regression coefficient.  The latter  spans a short interval [0.962; 0.984]  for the $N_{c,r}$  plans, but a slightly  larger one  [0.931; 0.983] for $N_{i,r}$.

 It is observed from Table \ref{table:parametersFRLav3} that the best fits do not always correspond to the expected range $m_2>0$ and $m_3>0$.   For  $N_{c,r}$, only the present administrative distribution, R22, or the B15 plan, have both $m_2$ and $m_3$ exponents positive. In contrast, all, except M11, have both exponents positive in the     $N_{i,r}$  cases. In the M11 case, one reason  beside these features, might be the small number of data points, - even though one often expects a greater fit precision in such a case. In fact, the M11 plan has $quasi$ the best $R^2$.  Nevertheless, the $N_{i,r}$   data for the M11  case markedly has no  sharp decay at high rank, - since the Corse data point is missing, because the island is merged with another region in M11.  Interestingly,  this effect  ($m_3\sim0$)  indicates a more equal number distribution in population between the 11 regions in this plan, except for the highly populated region  ($m_3\neq m_2$).
 
  Notice that the  mid range slope, i.e. $-(m_2+m_3)\; (r_M/2)^{(m_3-m_2-1})$,  for the fit function is an indication of the $non-equality$ distribution.  It is easy to derive that this slope can be = 0, i.e. $m_3=-m_2$, at $r/(r_M+1)= m_2/(m_2-m_3)$, whence requesting $m_3 > 0 > m_2$, for such an "equality condition" case.  Such cases do not exist, as can be deduced from Table \ref{table:parametersFRLav3}. In contrast, the largest  inequality corresponds to the largest sum $m_2+m_3$. Interestingly, - from a political view point discussion, the largest slopes occur for  the new P13 in the case of  $N_{c,r}$, but for the present R22 case for $N_{i,r}$.  Thus, the most recent plans do not appear quite convincing  from both city and population number distribution points of view.
  
  Thus, at this investigation level, it  can be concluded that  the present R22 case is  quite fine from a theory of settlements point of view.  The next best is B15. However, these considerations admit an inequality in distributions,- which might be contrary to criteria of equality (recall the motto of the French Republic!). One counter-argument might be based on the  point that the discussion does not pertain to equal  maximum size systems, thus compares different systems. A more "universal" view point should be taken! To  investigate this argument, a probability framework seems pertinent, as discussed in Sect. \ref{sec:entropy}.
 
\section{Entropy connexion}\label{sec:entropy}
 

One can consider to have  access to a sort of  "probability" for finding a certain "state"  (size occurrence)  at a certain rank, through
 \begin{equation} \label{pr}
p(r) =  \frac{N(r) }{\sum_1^{r_M} N(r) } \; \sim \;   \frac{y(r) }{\sum_1^{r_M} y(r) }Ê
\end{equation}
where $N(r)$ stands for  $N_{c,r}$  or $N_{i,r}$. 

From Eq.(\ref{pr}), one can obtain something which looks like the Theil index   \cite{Theil}
 in economy or  the Shannon entropy  \cite{shannon}
  in information theory,  $ S\equiv   -\sum_1^{r_M} p(r)\; ln (p(r))$. It 
has to be compared to the maximum disorder number,  i.e.   $ln (N)$. Whence we define the relative distance  to the maximum entropy as 
 \begin{equation} \label{d}
 d=1-\frac {S }{ln(N)}\;. \end{equation} 
A small $d$ value would indicate a state of full disorder (or "equal order"). Recall that non-equilibrium systems are those with  growth (... or decay !)  potential, i.e. those susceptible of evolution. Recall also  that the   notion of entropy maximization ($d \rightarrow  0$) of human systems is different from the concept of entropy increase in thermodynamics.

   Values of interest, $d_{c,r}$  or $d_{i,r}$,  in obvious notations, are reported in   Table \ref{table:dvalues}.  First of all, a good agreement is found between the  $N_{c,r}$  and $N_{i,r}$  cases, i.e.  $d_{c,r}$ and  $d_{i,r}$ smoothly decreases and increases respectively, as a  function of $N_r$,  accepting standard error  bars,  with the notable exception  for  the  M11 plan $d_{c,r}$   which has a very small value, $\sim$ that of the  present R22.     Notice that the present R22 map is that with the closest  distances to 0. The two smallest values occur for the M11 plan,  thus  suggesting that  it is not  drastically different from the R22, but still better from an entropic point of view.  The new P13 plan has, from  this  entropic point of view, very similar  $d_{c,r}$  and $d_{i,r}$ values, but they are not small. In fact, it has the largest $d_{c,r}$, indicating the most drastic changes  with respect to the present  time, from a city number  distribution point of view.


\section{Conclusions}\label{conclusions}

 Several questions were raised in the introduction,due to the multiple plans which have been proposed in order to reduce the number of regions in metropolitan France. One aim was to develop an objective analysis of the plans. Therefore, this paper provides a statistical analysis based on two numerical criteria: (i) the  distribution of  municipalities  between France regions, and (ii)  that of populations, according to those various plans proposed in recent  times.   These are standard variables in settlement analyses, because they rely on rather reliable numbers. It has been found that  several plans  weakly obeyed classical settlement models through their respective rank-size relationships. However, several cases are intriguing,  in not fulfilling theoretical behaviors. Considering an entropy (disorder $vs.$ order) criterion would seem to avoid a contempt argument about  this scientific approach.  From such an analysis point of view, it seems that the  city number or population number give  opposite optimization suggestions. In fact,  it is concluded that the M11 plan, based on the most  simple reduction plan,  seems the best for both criteria. The future P13 plan, on the contrary, seems in this respect far from respecting the  second term in the motto of the French Republic ({\it Libert\' e, Egalit\' e, Fraternit\'e}).
 
Next, the (fundamental ?) question is of philosophical nature: is disorder better or worse than order?  Often, order is claimed to be  more equalitarian. However, it is known that disorder is the source of growth, - quasi all systems of interest are non-equilibrium systems. Therefore, the "state of disorder"  is a criterion to be further examined. It would be interestingly monitored  for future data mining.

However, the evolution of a non-equilibrium systems, like a country state, is surely far from being described in terms of  a simple set of differential equations, - for a few variables. Furthermore,  a  long term growth  or decay  evolution is quasi  unpredictable, - the more so since the so called initial conditions are far from being precise. Moreover, there is no doubt that an objective set of criteria  has a weak weight with respect to short term political interests, - even hidden. It seems a little  bit strange also  that economic considerations have not been given much weight in any plan. One can admit that this would lead to {\it never ending} debates.
 

 {\it In fine}, notice that the final voted plan  refers to the  merging of regions, without questioning whether a re-distribution of departments  across previous region borders, or any gerrymandering, could  lead to an optimization in searching for the solution in the reduction of the number of regions problem.  Note that this will be possible up to 2019, but within a very strict political scenario  demanding several consensus with   a 3/5 majority.

\vskip0.2cm
 {\bf Acknowledgements}
This work has been performed in the framework of COST Action IS1104 "The EU
in the new economic complex geography: models, tools and policy evaluation". Moreover, this paper is part of scientific activities in COST Action TD1210 'Analyzing the dynamics of information and knowledge landscapes'.
 
\section{Appendix. Department analysis}\label{RPALav3}

  For completeness, the number distribution $N_{c,d}$ of municipalities  in the FR departments can be discussed within the main text framework.  Notice that  ranking this data is more sensitive to the merging or  creation of municipalities , since the numbers are smaller than in the region cases.   E.g., the department having the  greatest number of municipalities , 895, is Pas-de-Calais,  among the 101 departments. In contrast,  another example:  Paris Department is ranked 101, - in fact,   this department contains only one city, the capital of France, but belongs to the region Ile-de-France which has   about 1281   municipalities .  Observe  that Paris ranking changes from 101 to 96  under a  DOM-TOM "elimination  process".   The Paris department is clearly an outlier.  See some characteristics of the distributions in Table \ref{Tablestatcityperregiondept12-14FR}.

\begin{figure}
\includegraphics[height=7.0cm,width=7.8cm] 
{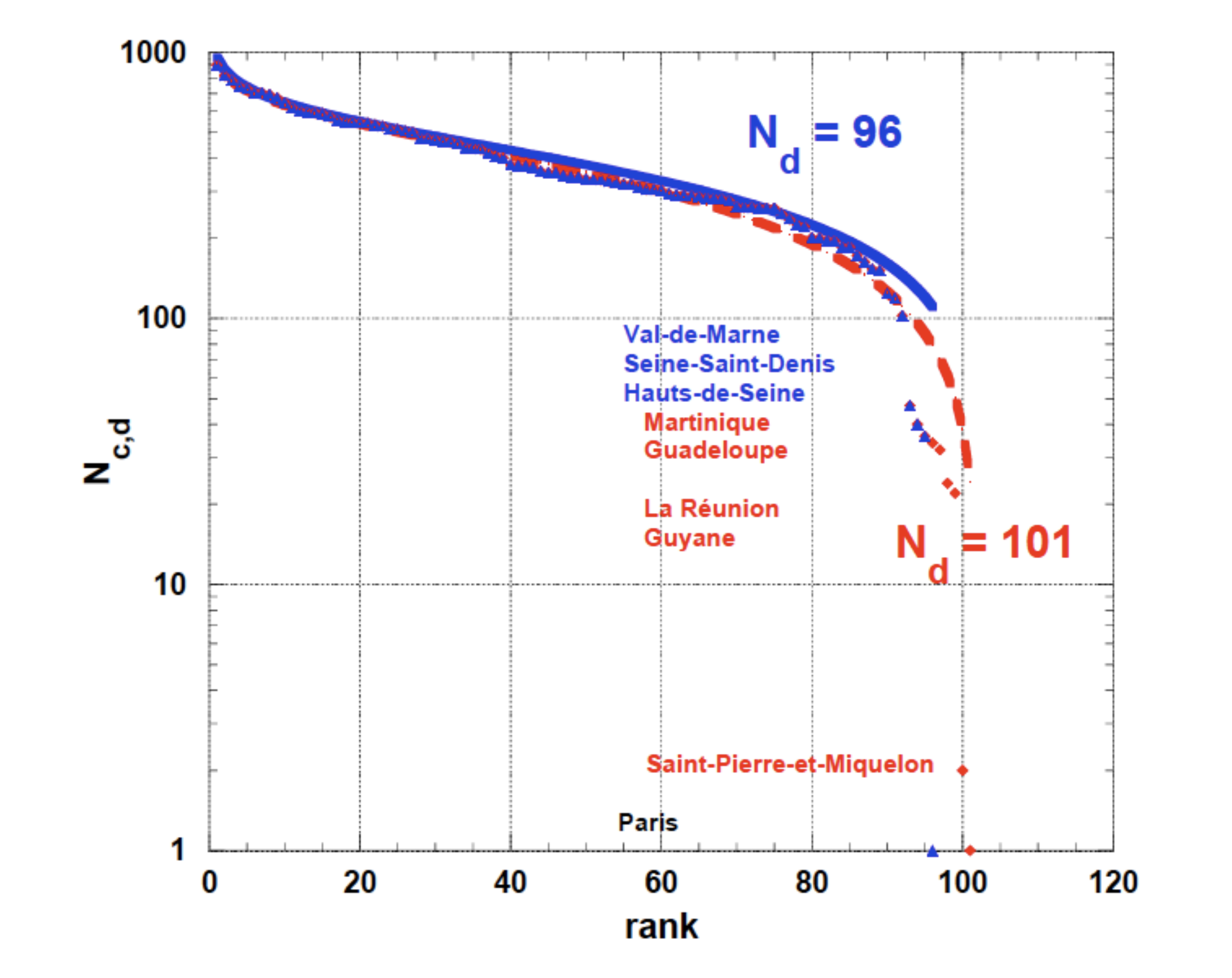}
 \caption   { Semi-log plot of the relationship between the number, $N_{c,d}$, of
municipalities  in the 101  FR departments (red diamonds) ranked by decreasing order of "importance" and in the 96 metropolitan departments (blue triangles); the best    3-parameter   function  fits values,  Eq. (\ref{Lav3}),
are shown. N.B. Paris changes rank, from 101 to 96;  best   3-parameter  fit values are
found in Table   \ref{TableLav3fitparamFRNcd101Ncd96}
.}
 \label{fig:Plot8FRNcd101Ncd96}
\end{figure}

For the demonstration, consider a semi-log plot of the number, $N_{c,d}$, of municipalities  in the (metropolitan)  96 and (all) 101  FR departments ranked by decreasing order of "importance" on Fig. \ref{fig:Plot8FRNcd101Ncd96}.  Interestingly, the behavior is similar to the $N_{c,r}$  cases.  However, the occurrence of outliers  cannot be debated.  They lead to very different fit parameter values. 
   The  best   3-parameter   function  fit values for $N_{c,d}$     
 are found in Table \ref{TableLav3fitparamFRNcd101Ncd96}.

 The DOM/TOM effect is greatly emphasized, pointing to a huge disparity between the mainland and the territorial communities. Fig. \ref{fig:Plot8FRNcd101Ncd96} allows to emphasize  the  (visually and numerically annoying) contribution from outliers in such fits and analyses.  




 




  \clearpage
 \newpage

 \clearpage
 \newpage
  \begin{table}
  \begin{tabular}{|c|c|c|c|c|c|c|c|c|c|c|c|c|c|}   \hline 
$N_{c,r}$&  R22&	B15& H14&P13&V12&M11 \\	\hline		
$A m_1$ 	&	915$\pm$120	&	638$\pm$148	&	211$\pm$ 91	&	216$\pm$114	&	601$\pm$241	& 1774$\pm$270  \\			
$m_2$ 	&	0.155$\pm$0.081	&	0.012$\pm$0.042	&	-0.101$\pm$0.068	&-0.099$\pm$0.085 &	-0.015$\pm$0.077&	0.108$\pm$0.036 \\
$m_3$ 	&	0.390$\pm$0.038 	&	0.674$\pm$0.078	& 1.162$\pm$0.153	& 1.227$\pm$0.195	& 0.864$\pm$0.151	&0.458$\pm$0.058   \\	 \hline	
R$^2$&  0.980&	0.973&0.968&0.962&	0.965&0.984 	\\	\hline	\hline	
$N_{i,r}$&  R22&	B15& H14&P13&V12&M11 	\\	\hline	
$A m_1/10^6$ 	&	7.36$\pm$1.97	&	5.66$\pm$2.81	&	7.4$\pm$2.0	&	4.2$\pm$1.8	& 43$\pm$1.9	&	13.96$\pm$2.2 	 \\			
$m_2$  &	0.678$\pm$0.034	&	0.521$\pm$0.083	&	0.510$\pm$0.070	&	0.382$\pm$0.080 &	0.334$\pm$0.092 &	0.554$\pm$0.042  \\		
$m_3$ 	&	0.141$\pm$0.084  & 0.269$\pm$0.175 & 0.168$\pm$0.131	&	0.385$\pm$0.157	&0.379$\pm$0.169	& -0.053$\pm$0.063   \\	 \hline	
R$^2$&  0.983&	0.936&0.948&0.947&	0.931&0.981 	\\	\hline 
\end{tabular}
\caption{ Parameter values allowing with Eq.(\ref{Lav3})  nice fits, as indicated by the regression coefficient   R$^2$  values.} 
\label{table:parametersFRLav3}
 \end{table}
 \begin{table}
 \hspace{-2cm}
 \begin{tabular}{|c|c|c|c|c|c|c|c|c|c } \hline 
$N_{c,r}$ &  R22&	B15& H14&P13&V12&M11  \\	\hline	
  $ln(N_r)$	&	3.0910 	&	2.7081	&	2.6391	&	2.5649 	&	2.4849 	&	 2.3979  	\\
$-\sum p\;ln(p)$ &3.0126	&	2.6056 	&	2.4809 	&	2.3871	&	2.3260	&	2.3334  \\ 
$d_{c,r}$&0.0254	&	0.0378 	&	0.0599 	&	0.0693 	&	0.0639 	&	0.0269 	\\
\hline \hline
$N_{i,r}$ &  R22&	B15& H14&P13&V12&M11  \\		
  \hline
   $ln(N_r)$	&	3.0910 	&	2.7081	&	2.6391	&	2.5649 	&	2.4849 	&	 2.3979  	\\
$-\sum p\;ln(p)$ &	2.8284 	&	2.5128	&	2.4837	&	2.4038	&	2.3419	&	2.3178  \\ 	
$d_{i,r}$&0.0850	&	0.0721 	&	0.0589 	&	0.0628 	&	0.0575 	&	0.0334 	\\\hline
\end{tabular}
\caption{Relative distance ($d_{c,r}$ and $d_{i,r}$) values, Eq.(\ref{d}),  to full disorder} 
\label{table:dvalues}
   \end{table}

\begin{table} \begin{center}
\begin{tabular}[t]{ccccccc}
  \hline 
 $r$&    Region name&$N_{c,r}$&$N_{d,r}$&$N_{c,r}$ \\
  &     & Jan. 2012& &Jan. 2014 \\ \hline
1	&	 Midi-Pyr\'en\'ees 	&	3020	&	8	&	3020	\\
2	&	 Rh\^one-Alpes 	&	2879	&	8	&	2874	\\
3	&	 Lorraine 	&	2339	&	4	&	2338	\\
4	&	 Aquitaine 	&	2296	&	5	&	2296	\\
5	&	 Picardie 	&	2291	&	3	&	2291	\\
6	&	 Bourgogne 	&	2046	&	4	&	2046	\\
7	&	 Champagne-Ardenne  
&	1954	&	4	&	1953	\\
8	&	 Centre 	&	1841	&	6	&	1841	\\
9	&	 Basse-Normandie 	&	1812	&	3	&	1812	\\
10	&	 Franche-Comt\'e 	&	1785	&	4	&	1785	\\
11	&	 Nord-Pas-de-Calais 	&	1545	&	2	&	1545	\\
12	&	 Languedoc-Roussillon 	&	1545	&	5	&	1545	\\
13	&	 Pays de la Loire 	&	1502	&	5	&	1496	\\
14	&	 Poitou-Charentes 	&	1462	&	4	&	1460	\\
15	&	 Haute-Normandie 	&	1419	&	2	&	1420	\\
16	&	 Auvergne 	&	1310	&	4	&	1310	\\
17	&	 \^Ile-de-France 	&	1281	&	8	&	1281	\\
18	&	 Bretagne 	&	1270	&	4	&	1270	\\
19	&	 Provence-Alpes-C\^ote d'Azur
	&	963	&	6	&	958	\\
20	&	 Alsace 	&	904	&	2	&	904	\\
21	&	 Limousin 	&	747	&	3	&	747	\\
22	&	 Corse 	&	360	&	2	&	360	\\ \hline
&	 SUBTOTAL 	&36 571	 	&	96  & 36 552	\\\hline
23	&	 Martinique 	&	34	&	1	&	34	\\
24	&	 Guadeloupe 	&	32	&	1	&	32	\\
25	&	La R\' eunion	&	24	&	1	&	24	\\
26	&	 Guyane 	&	22	&	1	&	22	\\
27	&	 Mayotte 	&	17	&	1	&	17	 \\\hline
	&	 TOTAL 	&	36 700	&	101 &36 681	\\\hline
\end{tabular}
\caption{Number $N_c$ of  municipalities   and of departments ($N_d$),  
 in  the (22  Mainland and Corse + 5 DOM-TOM)  France  regions,  on Jan. 01,  2012 and 2014.
 }  \label{TableNcityperregionFR}
\end{center} \end{table}

 \begin{table} \begin{center}
     \begin{tabular}{|c|cc|cc|cc|}
\hline  
&  \multicolumn{2 }{|c|}{$N_{c,r}  $ }  &   \multicolumn{2 }{|c|}{$N_{c,r} $  }  &  \multicolumn{2 }{|c|}{$N_{c,d }$}    \\ 
&  \multicolumn{2 }{|c|}{  Jan.01, 2012   }  &   \multicolumn{2 }{|c|}{  Jan.01, 2014    } & \multicolumn{2 }{|c|}{  Jan.01, 2014   }  \\
 \hline
min 	&	17	&	360	&	17	&	360	& 1 (**) &	 1 	 (**)\\
Max 	&	3020	&	3020	&	3020	&	3020	& 895	&	 	895	\\
$N_c$ &	36 700	&	36 571	&	36 681	&	36 552 	&	36 681	&	 	3 652	\\
$N_r$ or $N_d$	&	27	&	22	&	27	&	22	& 101&96	\\
Mean ($\mu$)	&	1359.3	&	1662.3	&	1358.6	&	1661.5 	&	 	363.18&	380.75	\\
Median ($m$)	&	1462	&	1545	&	1460	&	1545&	 	332		& 339.50	\\
RMS	&	1608.4	&	1781.8	&	1607.6	&	1780.9&	 	413.32	& 	423.91		\\
Std Dev  ($\sigma$)	&	876.30	&	656.61	&	875.94	&	656.44 	&	198.31		&	187.33\\
Std Err 	&	168.64	&	139.99	&	168.57	&	139.95 	&	 19.732	&	19.119	\\
Skewness	&	-0.1017 	&	0.2179 	&	-0.1018 	&	0.2167	&	 0.3331	& 	0.4416 	\\
Kurtosis	&	-0.7728	&	-0.2192	&	-0.7745	&	-0.2232	&	 	-0.2857	&	-0.1904 	\\
\hline
 $\mu/\sigma$ &1.5512&2.5316&1.5510&2.5311 &1.8314& 2.0325 \\
$3(\mu-m)/\sigma$&-0.3516&0.5359&-0.3473& 0.5324&0.4717&0.6606      \\    \hline
\end{tabular}
\caption{Summary of   statistical characteristics for the
number distribution of municipalities   ($N_c$)  in the various regions   ($N_{c,r}$) or departments ($N_{c,d}$)  in  FR on different years.  (**) N.B. Paris forms   a department  with only 1 city, itself. }
\label{Tablestatcityperregiondept12-14FR}
\end{center} \end{table}

   \begin{table} \begin{center}
     \begin{tabular}{|c|cc|cc|c}
\hline & $N_{c,d}$&$N_{c,r}$& $N_{c,d}$&$N_{c,r}$    \\
  & \multicolumn{2 }{|c|}{$whole$ FR} &  \multicolumn{2 }{|c|}{ FR$_{metrop}$}  \\ 
  \hline
 $A\;m_1$& 446.4&111.0& 848.0	&916.1     \\ 
 $m_2$ &0.131&0.048&0.148&0.155 	 \\
 $m_3$  & 0.654&0.991&0.525&0.389    \\\hline
    $R^2$  & 0.989&0.955&0.990 &0.978  	   \\\hline
\end{tabular}
\caption{  Best fit parameter values  (top) in Eq. (\ref{Lav3})    for the $N_{c,d}$ data, distinguishing 101 or 96 departments; see Fig.\ref{fig:Plot8FRNcd101Ncd96}; some $N_{c,r}$   data fit values are repeated for  ease.} \label{TableLav3fitparamFRNcd101Ncd96}
\end{center} \end{table}
\end{document}